# Robotics: Science preceding science fiction

Hortense Le Ferrand

Hortense Le Ferrand, School of Mechanical and Aerospace Engineering; and School of Materials Science and Engineering, Nanyang Technological University, Singapore; hortense@ntu.edu.sg

Robots and artificial machines have been captivating the public for centuries, depicted first as threats to humanity, then as subordinates and helpers. In the last decade, the booming exposure of humans to robots has fostered an increasing interest in soft robotics. By empowering robots with new physical properties, autonomous actuation, and sensing mechanisms, soft robots are making increasing impacts on areas such as health and medicine. At the same time, the public sympathy to robots is increasing. However, there is still a great need for innovation to push robotics towards more diverse applications. To overcome the major limitation of soft robots, which lies in their softness, strategies are being explored to combine the capabilities of soft robots with the performance of hard metallic ones by using composite materials in their structures. After reviewing the major specificities of hard and soft robots, paths to improve actuation speed, stress generation, self-sensing and actuation will be proposed. Innovations in controlling systems, modeling, and simulation that will be required to use composite materials in robotics will be discussed. Finally, based on recently developed examples, the elements needed to progress toward a new form of artificial life will be described.

## Introduction: A shift in vision in robotics

Well before robotics became a scientific research field in its own right, synthetic machines that could live alongside humans were present in our imaginations (**Figure 1**). The myths of the Pygmalion or the Golem[1] in Greek and Jewish folklores already mention human-like creatures arising from magic or the power of the gods. Since the industrial revolution and the resulting spread of the metal industry, metallic machines have inspired utopic worlds with self-driving transportation such as the steam Elephant from Jules Verne[2] or the rule of conscious machines in Metropolis.[3] With a Western European view—which might depart from Eastern cultures[1,4]—it is noticeable that as fictitious robots gain in autonomy, they start threatening human society. The Maschinenmensch from Fritz Lang, a beautiful—yet heartless—metallic robot, drives the rebellion of workers to ruin Metropolis[3]; Cybernauts,[5] or space robots[6] are used as killing machines; and the Terminator is sent to kill and terminate humanity.[7] In contrast, the goal of developing technological robots in industry is to help society and improve the lives of human beings. Low-skilled and repetitive manual labor has slowly been replaced by machines in manufacturing plants;[8] smart machines[9] and exoskeletons[10] have been developed to accelerate the rehabilitation of injured patients; and home robots have taken over cleaning tasks.[11] Beyond substituting for humans in low-skill and tedious tasks, robots have also been created to explore areas forbidden to humans, such as space[12] or the deep sea.[13] Lightweight and deformable robots that can interact closely with humans have emerged as soft robots at the end of the 20th century thanks to structural properties close to those found in nature. This is well illustrated by the booming number of publications in soft robotics.[14]

However, soft and kind robots in Western fiction are only slowly emerging. The most well-known is Baymax, featured by Walt Disney in 2014.[15] This shift from the threatening, gray, cold, and heartless metallic robots[13] toward friendly ones is coincident with the use of soft materials, characterized by conformability, colors, and constant adaptation to the environment. Bridges between material properties and our emotions have indeed been reported in several studies,[16,17] as supported by the use of soft robots in health care and medical applications, such as e-skins[18] and targeted drug delivery and surgery.[19] It is interesting to note that this opening of the mind came after advances in research and science.

However, despite the remarkable progress in soft robotics research, there is still a strong demand for innovation for more practical issues. Indeed, to explore a larger panel of applications, we still need to improve the motion capabilities of existing soft robots to enhance speed and control, to improve their resilience to environmental constraints such as temperature, flow, and accidental shocks, and their autonomy through programmable self-sensing and self-actuation. Furthermore, to be present in our production lines, cities, or homes, these soft robots need to be functional in diverse environments. Because of this, there is a need to increase the number of tasks that these robots can perform, to increase their resilience and lifetime, and to decrease energy consumption by increasing autonomy. One way to tackle this need could be to optimize the materials they are constructed from. With the development of new materials systems that offer advantages to both hard and soft robotics, new manufacturing methods, modeling, and robotic control strategies will also be required.

To complement the numerous excellent papers on the future of soft robotics,[20–23] this article will focus on the materials' scale strategies that could be implemented to transition from soft robots to stiffer composite robots. The unique features and advantages of soft robots and how they have revolutionized the field will first be described. Then, the current limitations in soft robotics will be discussed to determine the primary needs for innovation. Paths to create stiffer robots that combine the benefits of soft robotics with the performance of traditional hard metallic robots will be reviewed. Finally, a future of robotics will be suggested, taking into account the exponential pace at which research and scientific advances progress.

**The revolution of soft robotics**

Mainly based on organic materials, soft robotics has revolutionized the applications of robots by the creation of compliant devices that have multiple sensing capabilities; are directed or self-actuated; are able to interact and integrate with living systems; and are compatible with fast, customizable, and scalable fabrication techniques. Soft robots depart from traditional hard metallic robots by the materials from which they are constructed, the fabrication techniques, and their properties and applications (**Figure 2**).

Composed primarily of organic materials, soft robots are inherently soft, stretchable, and conformable. These mechanical properties are highly desirable for mimicking biological tissues and muscle actuation. For example, elastomeric vessels filled with air or liquid can inflate in predictable ways under a rise in pressure[32] or an electrically directed fluid flow.[33] These soft actuators can lift weights of up to 20 kg, sustain more than 100,000 cycles, and change shape to fold specifically around a fragile egg or fruit.[33–35] In the absence of corrosion-sensitive materials, these artificial muscles can be functional underwater and rendered transparent using materials with matching refractive indexes.[33,34]

Furthermore, soft matter and, in particular, hydrogel-based systems, allow an actuation that does not require power from a battery or motor system located on board. From this perspective, soft matter can be self-actuated and therefore has increased autonomy. For example, small variations in environmental conditions such as pH, temperature, or hydration levels can trigger large changes in volume.[36,37] If the hydrogel structure is anisotropic in swelling properties, a change in shape can accompany the change in environmental conditions. This strategy has been applied to create self-morphing actuators that can delicately grab fragile objects.[33,37,38]

In addition, the large chemical diversity of organics offers the potential to couple flexible mechanical properties with numerous functionalities such as transparency,[39] self-healing,[38] biocompatibility,[40] and electrical conductivity.[41,42] These properties can be incorporated in the soft robot thanks to multimaterial manufacturing paths. In particular, additive manufacturing and three-dimensional (3D) printing of polymeric materials allow the fabrication of customized elements with complex shapes and heterogeneously distributed compositions, at a quick and scalable printing pace.[43]

With the development of soft robotics, disruptive impacts have been made in fields where traditional metallic robots have hit limits: health and medicine, where interactions with soft biologic tissue required similarly soft mechanics, and in entertainment and care devices interacting on a daily basis with humans.[21,40]

The use of entirely soft materials and machines to replace or augment the traditional metallic ones is desirable for many reasons. First, organic materials are lightweight in comparison to metals and therefore need a lesser power of actuation. Second, conformability and diverse grabbing capabilities permit safer handling for any object shape. Finally, enabled functions such as self-healing or the ability of being 3D-printed are valuable assets for customized systems adapted to specific tasks. However, despite the new combinations of properties made possible by soft robots, strong limitations exist in terms of their performance that restrict applications to those previously described.

**Stiffening soft robots via composites**

To be fully functional in most common environments, soft robotic systems need structural protection against heat, cuts, and shocks. Furthermore, if untethered strategies are explored, most systems still rely on external power and actuation via batteries or connections through cables and Wi-Fi antennas.[44] To overcome these limitations and produce resilient and highly functional robots, there is a need to transpose soft robotic capabilities into stiffer systems. This can be achieved using composite materials (**Figure 3**).

The first limitation of soft robots is the limited level of stress and loading that they can generate. While soft robots have the capacity to lift up to 100× more than their weight,[35] industrial or rescue robots still require higher loading capacities. One of the strongest industrial robots on the market, M-2000 from FANUC Company, has a payload capacity of 2.3 tons. To overcome this issue, organic matrices could be reinforced with stiff inclusions such as carbon-based or ceramic-based fibers, particles, or nanotubes. These composite materials have Young's moduli up to two orders of magnitude greater than the strongest elastomers used in soft robots (Figure 3a). Stiff composite actuators constructed from thermoset matrices reinforced with carbon nanotubes,[57] long carbon fibers,[56] or ceramic microparticles[45] can generate actuation stresses from 10 to $10^3$ MPa, which is of the same order as traditional metallic hard robots[53–55] (Figure 3b).

In addition, the presence of reinforcement and the thermosetting nature of the matrix reduces the sensitivity to environmental conditions as compared with hydrogels. Indeed, thermosets can be rendered hydrophobic and less sensitive to moisture by the addition of micro-reinforcements that decrease their porosity and prevent diffusion of chemicals. Furthermore, most of these composites can still perform mechanically at elevated temperatures around 70–80°C[58] and can also be modified for UV and weathering resistance[59,60] (Figure 3a). Finally, the micro-reinforcements can create toughening mechanisms that can increase the resilience of the composite to external mechanical damage. Coupled with self-healing strategies, such as the use of microcapsules containing a healing or curing agent, some reinforced thermosetting composites exhibit approximately 80% healing efficiency and perform up to $5 \cdot 10^6$ loading cycles.[61,62]

**Transposition of soft properties in stiff composites via microstructuring**

The downside of this mechanical stiffness is that actuation and morphing possibilities become restricted as compared to softer materials. To recover these properties, several strategies have been explored that allow directed and self-actuation capabilities through careful design of the composite reinforced matrix architecture (**Figure 4**).

Self-actuation and the ability to respond autonomously to external stimuli is typically introduced into stiff reinforced composites by enabling the volume change of an organic matrix by anisotropic solvent impregnation, thermal expansion, or electric and magnetic properties[63,64] (Figure 4a, left). The construction of multilayer structures with dissimilar volume changes between each layer can lead to programmable reversible morphing in response to external stimuli (Figure 4a, right). Such shape-memory composites (SMC) have found applications in deployable structures for aerospace[65] and robotic actuators.[66] Along with self-actuation, direct triggering can also be implemented at the composite material's level by using specific material properties such as piezoelectricity or ferroelectricity (Figure 4b). For example, reinforcing a polymeric matrix with short fibers that display intrinsic shape-changing actuation has resulted in flexible ferroelectric composite fibers suitable for robotic systems.[67]

Along with sensing capabilities, morphing and actuation can be intentionally programmed (Figure 4c) through local composite designs. Directionality of the reinforcements is one convenient method to control the local stiffness (Figure 4c, top) and the direction of the volumetric change. Multilayers with controlled local stiffness have been used to create self-shaping objects and SMCs.[68] In composites in which magnetic microparticles are used as reinforcements and are distributed with predetermined orientations and positions, external magnetic fields (Figure 4c, bottom) have been used to control remotely the precise shaping of the robot and to drive its locomotion inside a phantom stomach.[69,70] Finally, internal stresses can be built within microstructured stiff composites to exploit nonlinear mechanisms such as buckling and multistability[45,71] (Figure 4d). Bistability in epoxy shells reinforced with long or short particles has been explored to achieve a combination of fast actuation and high actuation stresses (Figure 3b). In a typical example, a thin laminate consisting of two layers with perpendicular directions of reinforcement and thermal expansion is built and cured. During cooling from the curing to room temperature, stresses accumulate leading to a nonplanar morphology. If the geometric dimensions allow sufficient stresses, this morphology will correspond to one stable state, while the symmetric morphology will constitute the other stable state.[72] The flipping from one state to the other—the snap through—occurs quickly once the energy barrier between the two stable configurations is reached. Another example of buckling instability used in robotics are kirigami structures where, upon stretching, a thin sheet deforms and bends out of plane to anchor on the asperities of surfaces. With anisotropic frictional properties in the sheet, the contraction of an actuator placed on top will pull the structure forward.[44,73]

Building structures from reinforced and locally designed materials is a path to combine soft robotic capabilities (i.e., self-sensing and actuation) and conformability with hard metallic robot-like performance (i.e., quick response and large stress generation). The use of composite materials nevertheless poses other challenges in their fabrication and actuation control, demanding the development of new strategies in these areas.

**Stiff composite-based robotics**

The use of stiff, reinforced composites to build robots promises high stress generation and fast directed or autonomous actuation. The advantages of robotic systems based on these materials as compared with traditional hard robots will be the reduced costs in energy consumption, thanks to autonomous actuation, and in maintenance, because of resilient composites and the adapted mechanics and functions. Autonomy in a composite-based robot could be achieved by implementing sensing and actuation at the materials' level, where this requires innovation in order to achieve local and decentralized commands. Furthermore, fabricating a complex and multimodal robot using stiff composites will also demand innovation in its manufacturing and in the simulation and modeling of its macroscopic response.

Controlling stiff reinforced composite-based robots will require us to rethink the actuation paths of complex structures. Indeed, encoding the actuation response at the materials' level, direct sensing, computation, and actuation can be decentralized, without the need for channeling all inputs toward a computational brain as in traditional fully actuated structures (**Figure 5**). This new approach describes the robotic materials proposed by Correll et al.,[74] where all the controls are located locally within the structure. This strategy is particularly interesting to increase autonomy since local sensing and actuation in response to an external trigger will determine the global response of the structure. Similar to a reflex, this response can be fast and cost little in transportation of information and computational power.

Furthermore, one tremendous advantage of soft robotics over traditional metallic robot engineering is the possibility of using additive manufacturing alone to fabricate the robot. With the use of reinforced composites to increase the mechanics and the performance of soft robots, additive manufacturing is challenged. Three-dimensional printing has proven to be a convenient tool to control local stiffness and directions in reinforcement via effects of shear forces,[75] ultrasound[76] or external fields,[77,78] and chemical diversity.[79] However, these are limited in terms of reinforcement concentration due to the increase in viscosity and the difficulties in obtaining a homogeneous and flowing composite mixture, which in turn restricts the mechanical performance. Current alternatives for the fabrication of composite-based robots are to use prepreg (pre-impregnated) long fiber-reinforced epoxies which can be assembled manually in specific ways.[56] However, this process only allows flat shapes to be constructed, with little chemistry diversity. Another approach is to form a thick composite mixture and use external fields to orient microparticles in specific directions as the viscosity is decreased with temperature.[45] However, much is still to be done to realize high degrees of structural and compositional control in composite materials, and to create complex, composite-based robotic systems.[80]

Optimization of both the manufacture and control of composite-based robotic systems could lead to the design of robots that could not only replace humans in certain tasks, but also in applications where humans are underperforming, such as for rescue or exploration (**Figure 6**). In contrast to current systems, the autonomy of the robots, their multifunctionality and their mechanical resilience would allow them to sample objects of any shape and to adapt to the environment for longer service. If soft robotics was inspired by nature due to the softness of our bodies, a more comprehensive comparison lies in the composite nature of our bodies, where hard and soft elements are intimately mixed, such as our bones and muscles.

**Toward an artificial form of life?**

The ultimate autonomous robot is an artificial machine that is able to wander on its own in any environment, much like a living creature. To this aim, some energy generation has to be on board the robot, along with self-growing options. With recent advances in biotechnology and tissue engineering, such robots might come to life.

Indeed, along with the development of 3D printing for soft robotics, bioprinting has demonstrated the possibility of printing materials comprising living cells. With the appropriate delivery of nutrients, the cells embedded within the material can grow, differentiate, and replicate to colonize the entire material, and to synthesize the cues appropriate for their environment, such that they ultimately modify the material entirely.[81] This principle has a direct impact on bioengineering and biomedicine for tissue regeneration. Recent examples have also shown how the presence of living cells inside an artificial construct can be used to secrete chemical compounds or to degrade pollutants.[82] Also, recent studies are exploring how living cells can be directly used as energy providers and actuators in biosyncretic robots.[83–86] The examples developed are still focusing on the soft mechanical range. However, given that hard and stiff materials are present in biology, it can be expected that similar results could be achieved in composite robotic systems.

If the use of living cells is exciting to give life to artificial robots, mimicking natural life via synthetic means might be even more desirable. Indeed, this would allow greater control and provide greater insight into the mechanisms by which the robot operates. The result would be a more rapid implementation of the strategy in specific applications and the exploration of properties and capabilities that go beyond those of natural organisms, which are the initial goals of robotics. With this in mind, plant-inspired growing robots have been developed based on soft technology—pneumatics inflation[87] or 3D printing of material.[14] The second strategy, where the robot consists of a 3D printer head that deposits material as the robot grows is potentially applicable to composite robots. An efficient self-growing robot could be imagined in the following way: a central unit would localize the presence of the materials necessary for its growth. After moving toward this source, the material is extracted from the environment, processed by the robot to make it ready for 3D printing, then printed in the direction to grow, and with the relevant properties as required for the robot to move forward or perform a task.

Finally, one can question the need of such self-growing robots or robotic forms of life. As stressed earlier, the interest in robotics is not to replace nature and humans but rather to be used in areas that are dangerous or undesirable to us: exploration of unknown environments, rescue and entertainment. As scientific research advances at a greater pace than science fiction, the development of these systems opens up many possible applications that have not yet been predicted.

**Conclusion**

Soft robotics has pushed the traditional field of hard robotics one step forward by allowing complex morphing, combinations of directed and autonomous sensing and actuating, and fabrication via 3D printing. To access a larger range of applications and to further improve their performance, current soft robots need mechanical resilience. Composite systems therefore appear an obvious choice, where they can potentially combine the advantages of soft and hard robotics. The vision for composite robotic systems is to create a robot that is fast and strong, that is autonomous, and which is adaptable and capable of complex morphing, as are living vertebrates. Finally, such robots could also be made to grow using living cells or other synthetic approaches. The examples of composite materials and composite robots discussed in this article highlight the challenges they pose, in terms of manufacturing, control, and modeling. If science fiction seems to fail at predicting the future of robotics, one can expect that the synergetic effort from chemists, material scientists, roboticists, engineers, and programmers, could cement this long-standing dream of autonomous synthetic machines.

**Acknowledgments**

Thanks to the Jiang Family Foundation and the MTI Corporation for their generosity. Thanks also to A. Rafsanjani for input on actuated soft structures and Q.-C. Pham and L.G. Zhan for review and feedback.

**Figure**

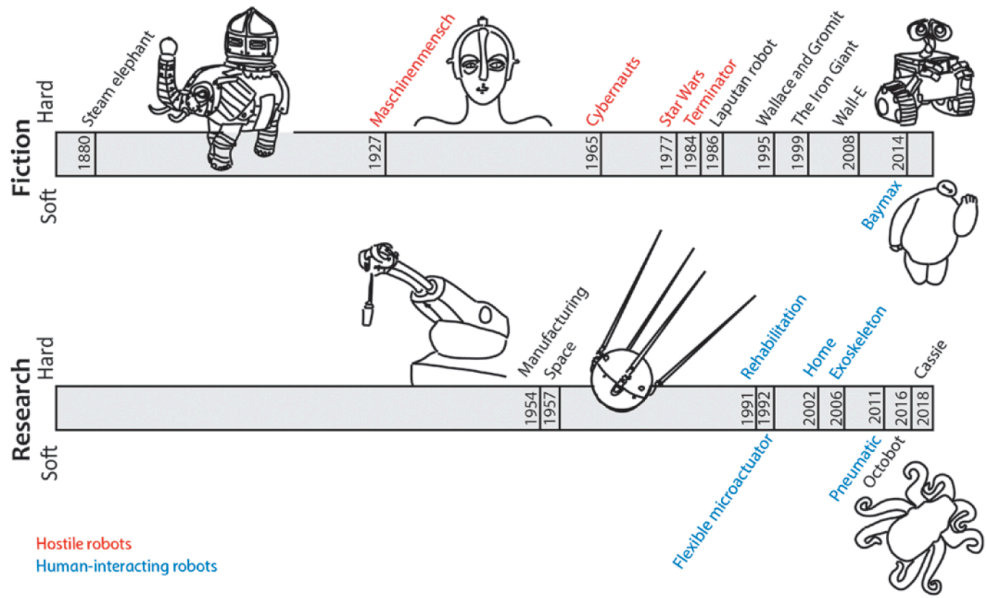

**Figure 1.** Timeline appearance of selected hard and soft robotic machines, both in fiction literature or cinema and in research, also highlighting a shift between the vision of robots as hostile machines (in red) in fiction in opposition to friendly and useful human-interacting tools developed in research, but slowly appearing in recent fiction work too (blue). Table I lists the references of the selected examples provided here.

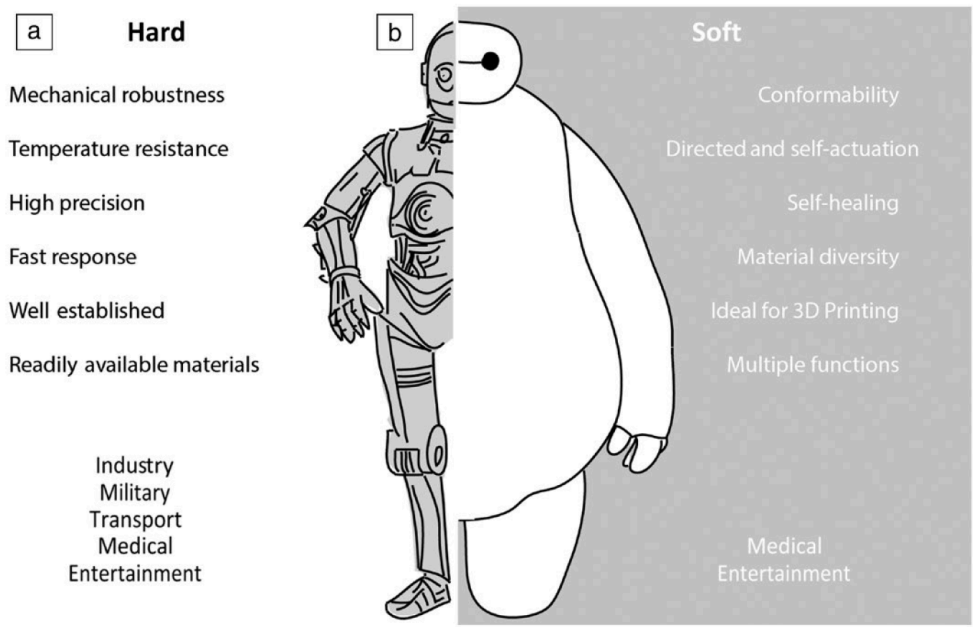

**Figure 2.** Selected properties and applications (capital letters) of traditional hard robots versus soft robots. The cartoons represent (a) a famous metallic humanoid[6] and (b) a soft character.[15]

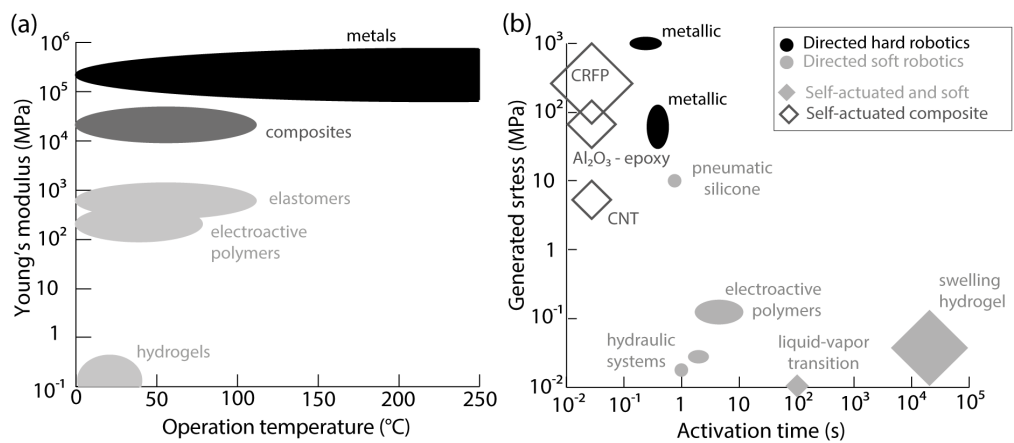

**Figure 3.** Comparison of the performance of hard, soft, and composite robotic systems. (a) Ashby-like plot representing the Young's moduli as a function of the temperature of operation of common materials used in robotic systems: metals (black), epoxy-reinforced composites[45,46] (dark gray) and polymers (light gray), such as elastomers,[29,47–49] electroactive polymers[50] and hydrogels.[39,51,52] (b) Ashby-like plot representing the generated stress from morphing structures as a function of their actuation time for directly-actuated hard metallic robotic systems,[53–55] self-actuated stiff composite robotics,[45,56,57] and directly and self-actuated soft actuators.[29,39,48,56,57] CNT = carbon nanotubes; CRFP = carbon fibre reinforced polymer.

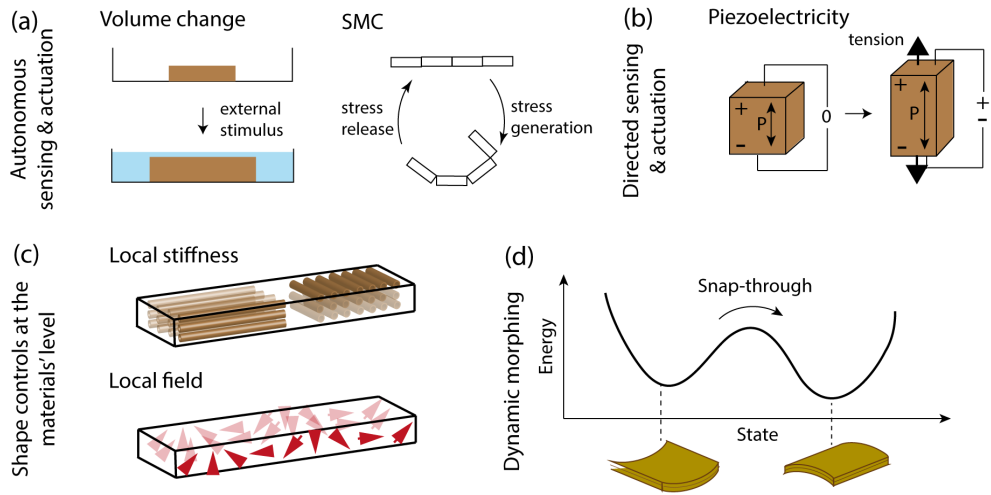

**Figure 4.** Example of strategies to embed actuation at the materials level in composite systems: (a) autonomous sensing, with (left) volume change in response to an external stimulus, and (right) shape-memory composites (SMC); (b) directed sensing and actuation through materials' properties such as piezoelectricity; (c) control of the morphing through the internal design of the material by (top) structuring with locally varying stiffness and (bottom) local properties; (d) dynamic morphing response by making use of mechanical instabilities such as bistability.

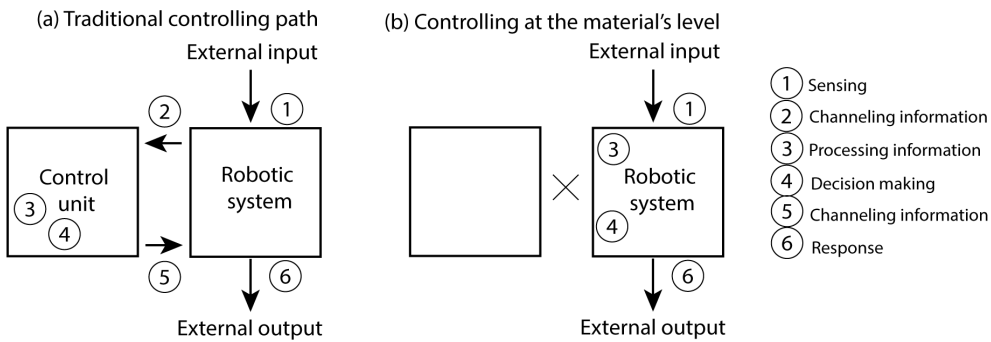

**Figure 5.** Controlling paths in robotic systems: (a) in traditional paths and (b) in future stiff composite robots, where the controls are decentralized at the material's level. The blank square to in (b) indicates the absence of a control unit.

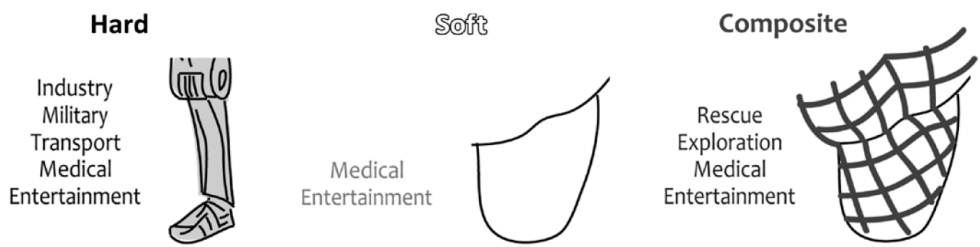

**Figure 6.** Expansion of the applications fields of robotic systems from hard to soft to composite.